\documentclass[10pt, conference, compsocconf]{IEEEtran}
\usepackage{times}
\usepackage[nolist]{acronym}
\usepackage{multirow}
\usepackage{verbatim}
\usepackage{cprotect}
\usepackage{graphicx}
\usepackage{url}
\usepackage[skip=1pt,small]{caption}
\usepackage{subfigure}
\usepackage{amsmath}
\usepackage{cite}
\usepackage{floatrow}
\usepackage{makecell}
\usepackage{boldline}

\newfloatcommand{capbtabbox}{table}[][\FBwidth]

\begin{document}
%
% paper title
% can use linebreaks \\ within to get better formatting as desired
\title{Scalable and Efficient Construction of Suffix Array with MapReduce and In-Memory Data Store System}

% author names and affiliations
% use a multiple column layout for up to two different
% affiliations

\author{\IEEEauthorblockN{Hsiang-Huang Wu*, Chien-Min Wang, Hsuan-Chi Kuo*, Wei-Chun Chung and Jan-Ming Ho}
\IEEEauthorblockA{Institute of Information Science, Academia Sinica\\
Taipei, Taiwan\\
\texttt{\{cmwang,wcchung,hoho\}@iis.sinica.edu.tw \{virtuoso.wu,hsuanchikuo\}@gmail.com}*}
}

% make the title area
\maketitle

\begin{abstract}
\ac{sa} is a cardinal data structure in many pattern matching
applications, including data compression, plagiarism detection and
sequence alignment.  However, as the volumes of data increase abruptly,
the construction of \ac{sa} is not amenable to the current large-scale
data processing frameworks anymore due to its intrinsic proliferation of
suffixes during the construction.  That is, ameliorating the performance
by just adding the resources to the frameworks becomes less cost-effective, even having
the severe diminishing returns.  At issue now is whether we can permit \ac{sa} 
construction to be more {\bf scalable} and {\bf efficient} for the everlasting accretion of data
by creating a radical shift in perspective.
Regarding TeraSort~\cite{owen2008x1} as our baseline,
we first demonstrate the fragile scalability of TeraSort and investigate what causes it	through
the experiments on the sequence alignment of a grouper (i.e., the \ac{sa} construction used in bioinformatics).
As such, we propose a scheme that amalgamates the distributed key-value store system
into MapReduce to leverage the in-memory queries about suffixes.
Rather than handling the
communication of suffixes, MapReduce is in charge of the communication of their indexes,
which means better capacity for more data. 
It significantly abates the required disk space for constructing \ac{sa} and better utilizes the memory,
which in turn improves the scalability radically.
We also examine the efficiency of our scheme in terms of memory and show it outperforms TeraSort.
At last, our scheme can complete the pair-end sequencing and alignment with two input files without
any degradation on scalability, and can accommodate the suffixes of nearly 6.7 TB in a small cluster composed of 16 nodes
and Gigabit Ethernet without any compression.
%Furthermore, neither MapReduce nor the distributed key-value store system is required to be modified
%for our scheme.

\end{abstract}


% For peer review papers, you can put extra information on the cover
% page as needed:
% \ifCLASSOPTIONpeerreview
% \begin{center} \bfseries EDICS Category: 3-BBND \end{center}
% \fi
%
% For peerreview papers, this IEEEtran command inserts a page break and
% creates the second title. It will be ignored for other modes.
\IEEEpeerreviewmaketitle

\section{Introduction}
\label{sec:intr}
%suffix array
Suffix Array (\ac{sa}), proposed in~\cite{manb1990x1} and enhanced in~\cite{abou2004x1}, is more widely used because of better
locality of memory reference and consumes less space than suffix tree~\cite{ohle2010x1}.
Table~\ref{tbl:example_sa} illustrates how to construct the \ac{sa} of \verb#SINICA$# in lexicographically ascending order, where
\verb#$# is a delimiter and lexicographically smaller than the other characters.
All possible suffixes are listed in the rightmost column and we sort them by comparing the characters from left to right.
On the other hand, \ac{sa}[$i$] is a sorted array composed of the indexes that indicate the corresponding sorted suffixes.
Suppose the length of a string is $n$ and the sorting algorithm is comparison-based.
The total number of suffixes and the time complexity would be $n$ and $O(n^2$$\log$$n)$ respectively
since we include the time spent on comparing the characters (i.e. $O(n$$\cdot$$n$$\log$$n)$).
Hence the demands for the fast construction of \ac{sa} has led to the development of linear
time algorithms through exploiting the characteristic of the suffixes~\cite{kark2003x1,kim2003x1,ko2003x1}.
Also, \ac{sa} construction that needs only $O(n)$ working space is presented in~\cite{hon2003x1}.

\begin{table}[htbp]
\cprotect\caption{Suffix Array of \verb#SINICA$#.}
\label{tbl:example_sa}
\footnotesize
\center
\begin{tabular}{|c|c|l|l|}\hline
Index $i$ &  SA[$i$]   & Sorted Suffix  &  Suffix   \\ \hline
0         &    6  &   \verb#$#                &  \verb#SINICA$# \\\hline
1         &    5  &   \verb#A$#               &  \verb#INICA$#  \\ \hline
2         &    4  &   \verb#CA$#              &  \verb#NICA$#   \\\hline
3         &    3  &   \verb#ICA$#             &  \verb#ICA$#    \\\hline
4         &    1  &   \verb#INICA$#           &  \verb#CA$#     \\\hline
5         &    2  &   \verb#NICA$#            &  \verb#A$#      \\\hline
6         &    0  &   \verb#SINICA$#          &  \verb#$#       \\\hline
\end{tabular}
\end{table}

%why not signle machine but distributed comuting
In the era of ''Big Data'', \ac{sa} construction confronting large volumes of data becomes very critical
and needs to be handled by the distributed computing frameworks such as Hadoop~\cite{hadoop}.
This is because the resources of the single machine, especially the memory, cannot afford these
copious suffixes derived from ''Big Data''.  More concretely,
\verb#libdivsufsort#~\cite{mori}, state-of-the-art \ac{sa} construction algorithm for single thread,
claims $O(n$$\log$$n)$ worst-case time using only $5n+O(1)$ bytes of memory space.
Say, there are $10^9$ strings and each of them consists of 20 characters (i.e. $n=20$).
If we want to perform in-memory sorting, the required memory would be 100 GB at least.
It implies switching from scale-up to scale-out is inevitable, but how good is scale-out?

%issues about just scale-out
Many distributed systems that emerge from behind
cloud computing improve speedup by the promise that cloud computing provisions unlimited resources.
This naturally leads to the focus on how we fit the application into a distributed computing framework
because the speedup is thought to be improved directly by scale-out: adding more resources to the systems.
We argue that such plunge into scale-out, inherently associated with the mechanisms of the distributed computing frameworks,
is not the panacea for speedup; instead, this desirable performance gain might
belie its poor capacity for the coming larger volumes of data.

%scale-out concerning scalability
We use scalability and efficiency as our criteria for how good the capacity of a distributed system for \ac{sa} construction is.
The outline is described in the following.
Two directions of scalability---$scalability_{1}$ and $scalability_{2}$---defined in~\cite{wein2006x1} are
our conceptual and main considerations.
To elaborate on $scalability_{1}$, we introduce three types of scalability from~\cite{bond2000x1} to explore
it in more detail aspects, and for $scalability_{2}$, we introduce the efficiency to understand whether
the incorporation of the in-memory data store systems into \ac{mr} is cost-effective or not.

%\begin{itemize}[topsep=-0.25ex,itemsep=-0.5ex]
\begin{itemize}
  \item $Scalability_{1}$ is the ability to handle increased workload (without adding resources to a system).
  \begin{itemize}
    \item Load scalability
    \item Structure scalability
    \item Space scalability
  \end{itemize}
  \item $Scalability_{2}$ is the ability to handle increased workload by repeatedly applying a cost-effective strategy for extending a system's capacity.
  \begin{itemize}
  %\item The ease of adding the resources
    %\item Space scalability
    \item Efficiency
  \end{itemize}
\end{itemize}

%efficiency
%On the other hand, to assess the efficiency of a system is not quite easy, especilly a distributed system.
%Conventioanally, the efficiency of a parallel system that is compute-intensive can be estimated by
%$\frac{T_s}{p \times T_p}$ or $\frac{speedup}{p}$,
%where $p$, $T_p$ and $T_s$ are the number of processors, the execution time on one of $p$ processors,
%the execution time on single processor respectively.  

%data store footprint model
Involving network, I/O, hierarchical storage, and computing units,
a distributed system nowadays cannot ascribe its performance gain to one kind of resources.
Thus, some measurement commensurate with time, for example latency or throughput, is necessary
to reason about the performance while being analyzed to get insights into which factor the performance is
bounded by.
We identify data store footprint with time, where the data store refers
to memory and disk in this paper.  The performance analysis of the data store footprint
is carried out by investigating which one of the four factors---CPU, memory, disk and network---is
bound to be the bottleneck mostly.

It is because reading and writing the
sheer size of data through I/O almost predominates in large-scale data processing that
the extent of space required can reflect the extent of time consumed.
We keep three observations concerned with data store footprint in perspective:

\begin{itemize}
  \item \textit{Fine-grained data movement}. Suffixes are tiny in comparison with the scale of the input.
  \item \textit{Data access pattern}.  Access of suffixes while sorting is irregular and very frequent.
  \item  \textit{Trading disk I/O for memory I/O}.  Not only does local memory access outperforms local disk access
  but also remote memory access competes it in latency~\cite{dean2009x1}.
\end{itemize}

%example that scale-out doesn't work and its efficiency is not satisfied
In~\cite{meno2011x1}, the \ac{sa} construction purely by \ac{mr} is proposed and
the experimental results show non-linear speedup within a range of 30-, 60- and 120-core cluster while the
problem size is fixed.
According to the authors' investigation, it is caused by writing the replications and not used values
(unsorted suffixes) to \ac{hdfs}.  Furthermore, we assess the performance improvement in terms of efficiency
and assume that the experimental result of the 30-core cluster is seen
as the baseline.  By $\frac{speedup}{p}$, the efficiency of the 60-core cluster and the 120-core cluster
are $\frac{1.45}{2} = 72.5\%$ and $\frac{1.53}{4} = 38.25\%$.
Despite the rough calculation, we still can learn that scale-out without discerning what kind of resource
matters might thwart the advancement in efficiency.  Regarding $scalability_1$, \cite{meno2011x1} claims the \ac{sa} construction
scales linearly.  Through the metrics of the experiments, we make deductions about its $scalability_2$
and find it scales linearly as well.  Nevertheless, we will prove that \ac{sa} constructed only by \ac{mr}
overloads the disks, thereby easily causing a breakdown in the scalability.
On the other hand, inspired by this analysis that \ac{mr} can scale linearly,
we delineate our scheme in a way that adopts \ac{mr} as the foundation and exhibits the keen insights of scalability
and efficiency with respect to $scalability_{1,2}$. 
Through the analysis of data store footprint, we show how our scheme can surmount the fragile scalability.

%experiment
% reads
% contrast to meno2011x1 one genome
It is always convincing to demonstrate how our scheme reaches the goals using a real application.
Sequence alignment is a very important application in bioinformatics, and highly
relies on two index structures---\ac{sa} and \ac{bwt}~\cite{meno2011x1}.  The latter can be
derived from the former.  As an illustrative example of our scheme, we use the authentic sequencing data of grouper genome.
The total input size is 64 GB (325,718,730 reads) and the length of each read is about 200 bp (i.e. 200 characters).
Without getting rid of any suffix,
the total suffixes including their indexes would be around 6.7 TB, hundred times the input size.
Furthermore, we start a \ac{sa} construction with TeraSort as
our baseline for analysis.  According to the deficiency identified in TeraSort, we develop our scheme
that evolves into a system that makes available more scalable and efficient \ac{sa} construction.
More specifically, there are 6.7 TB suffixes during the construction in our cluster composed of 16 nodes
and Gigabit Ethernet, and without sacrificing the speedup, our experiment takes
11 hours to generate the output that contains the suffixes and the indexes of the corresponding reads.
%in corresponding format for the following steps of sequence alignment.

\section{MapReduce}
\label{sec:mr}

\begin{figure}[htbp]
\centering
\includegraphics[scale=0.15]{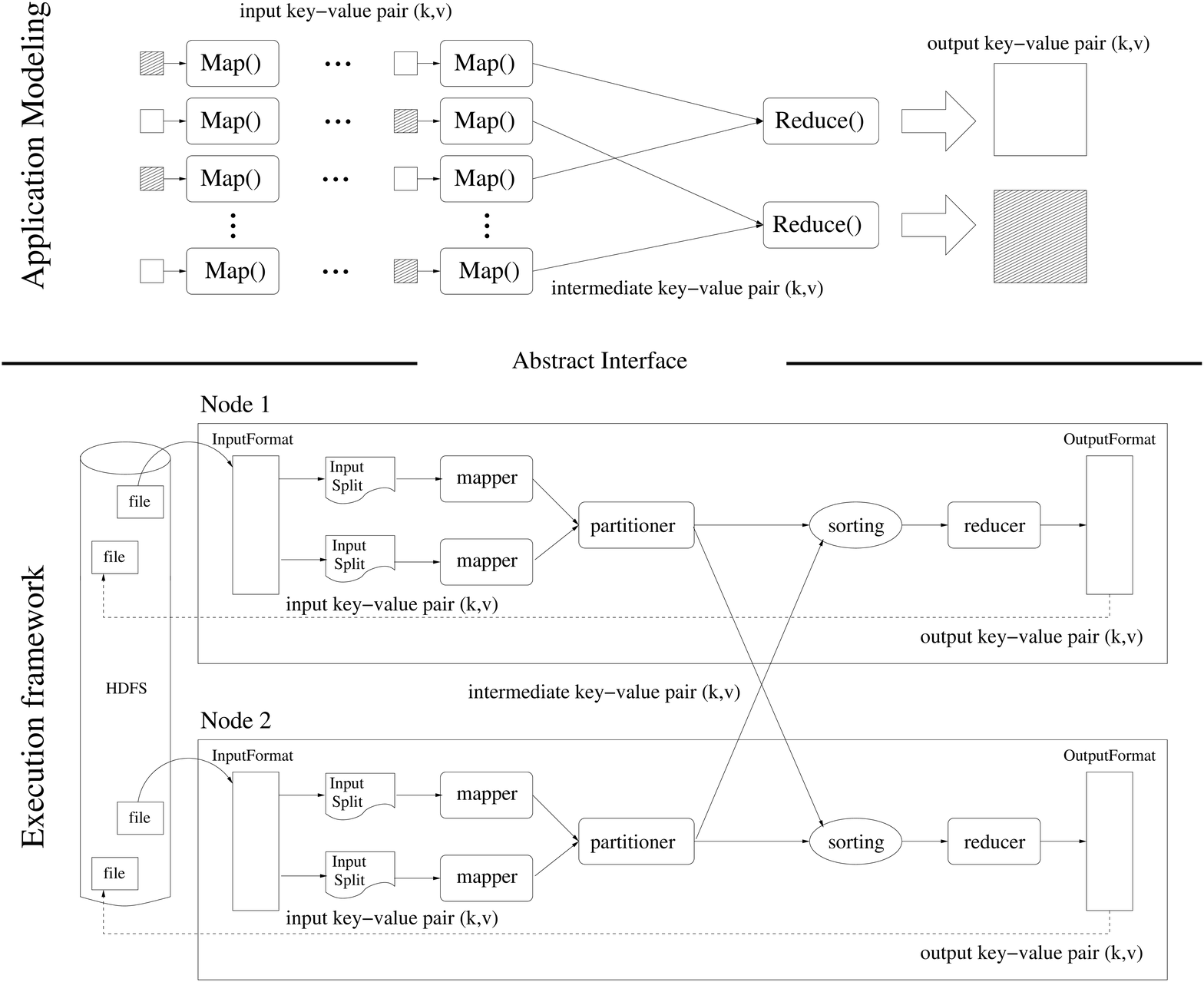}
\caption{The interpretation on application modeling and execution framework of \ac{mr}.}
\label{fig:mapreduce}
\end{figure}

\ac{mr} is the enabling technology for large-scale batch processing~\cite{Dean2004x1}.
The dataflow {\bf Map-Sort-Shuffle-Merge-Reduce} mainly constitutes \ac{mr}.  Figure~\ref{fig:mapreduce} illustrates
\ac{mr} in a manner that considers \ac{mr} as a programming model.  For application modeling,
the relation with \verb#Map()# and \verb#Reduce()# conforms with  $Map(k_1,v_1)$ $\to$ $[(k_2,v_2)]$
and $Reduce(k_2,[v_2]) \to [(k_3,v_3)]$, where the parentheses $(...)$ and brackets $[...]$ denote a key-value pair
and a list respectively.  Furthermore, \verb#Map()# deals with only one key-value pair (i.e. atomic) every time and
then generates the intermediate key-value pairs, whereas \verb#Reduce()# aggregates these intermediate key-value pairs associated
with the same key.
As the example shown in Figure~\ref{fig:mapreduce}, an application collects two different
types of data, white and striped, and then, generates two corresponding collections.
Suppose there are 100 units of data that are either white (50\%) or striped (50\%) and scattered over the input file.

%characteristics of MapReduce
Two steps are in this application: identifying every data and collecting the data of the same type.
Giving every unit of data a unique number (e.g., 1 to 100), we regard $k_1$ as the serial number and $v_1$ as the corresponding unit of data.
Whenever \verb#Map()# reads the key-value pair $(k_1,v_1)$, it recognizes $v_1$ and then emits the intermediate key-value pairs $(w,v_1)$
($(s,v_1)$) if $v_1$ is white (striped). As to the \verb#Reduce()#, it is in charge of collecting $v_1$ according
to the key of the intermediate key-value pairs and finally, outputs both the 50 units of white and striped data.
In the application modeling, we think in the way that each key-value pair has its own \verb#Map()# and the intermediate key-value pairs with
the same key (one group) are processed by one \verb#Reduce()#.

%implementation of MapReduce
The execution framework of our example is assumed to have two nodes of which each can accommodate two
slots for the map tasks (i.e. mappers) and one slot for the reduce tasks (i.e. reducers).
This assumption is for Hadoop 1.2 for simplicity.  For Hadoop 2 YARN, slots are replaced with containers
which are not limited to mappers or reducers.
In reality, the mapper deals with a batch of data called Input Split that contains more than one key-value pair.
On the other hand, the reducer is allowed to handle more than one group under the constraint that all the
intermediate key-value pairs associated with the same key must be processed by the same reducer.  Once the application
implemented in \ac{mr} starts, \verb#InputFormat# reads the input file from \ac{hdfs} and divides it into
several Input Splits.  In our example, the number of Input Splits is 4 and each of them contains 25 key-value pairs
(invoking \verb#Map()# 25 times).  The number of Input Splits also determines the number of mappers, whereas the number of reducers
can be specified by the programmers through the system settings.  Every mapper processes the 25 key-value pairs
and generates the intermediate key-value pairs. Partitioner dispatches them to the reducers according to their corresponding partition numbers.
Sorting is performed on the files spilled by a mapper according to the partition number and then, key in order.
Merging is applied to the Map outputs (fetched from mappers) of a reducer and groups them into one single file
before the reducer begins.
It is possible that one reducer might have several different intermediate keys (invoking \verb#Reduce()# several times).
In the end, \verb#OutputFormat# writes back the results to \ac{hdfs}.

%environment of our cluster
In the following, we experiment on Hadoop 2.7.2 with the cluster in size of 16 physical nodes.
Table~\ref{tbl:scalability} illustrates the total hardware resources and the resources managed by YARN.
There are two types of Intel(R) Xeon(R) CPU and each node is equipped with two CPUs of the same type.
E5620 and E5-2620 are quad- and hex-core CPU that can provide 8 and 12 threads respectively.
We assign 2 GB and 8 GB memory to a mapper and a reducer of which the heapsize is
1 GB and 7 GB respectively.  The total number of VCores is 128 (the default value of VCores is 8 for each node).
%Since our cluster permits 160 containers (320/2) at maximum regarding the memory, we let each node
%donate 10 VCores (160 VCores in total) to match the maximum number of containers.
The replication factor is set to 1 to avoid excessive data writing to the disk.
Note that we assign 1 GB memory to \ac{am} and ask all nodes to provide this extra 1 GB
memory to prevent the memory slots for reducers from the occupancy of \ac{am}.
For instance, we actually ask each node to to donate 17 GB memory so that, at most, 8 mappers and 2 reducers
can run concurrently.  To make such 1 GB not distracting, we omit it in the following
sections.

\begin{table}[htbp]
\cprotect\caption{Summary of the resources in a Hadoop cluster with 16 physical nodes.  
                  The number of VCores and memory is donated by each node evenly.}
\label{tbl:scalability}
\footnotesize
\center
\begin{tabular}{|c|c|c|}\hline
\multicolumn{3}{|c|}{\bf Resources Managed by YARN} \\\hline
VCores               & Memory   & Disk\\\hline
128                  & 256 GB   & 28.24 TB\\\hline
\multicolumn{3}{|c|}{\bf Hardware Resources} \\\hline
CPU                  & Memory   & Disk\\\hline
%\multirow{2}{*}{E5620 2.40GHz (10 nodes)}  & \multirow{4}{*}{1352 GB} & 825 GB (4 node)\\
\multirow{2}{*}{E5620 2.40GHz (10 nodes)}  & \multirow{2}{*}{48 GB (5 nodes)}  & 825 GB (4 nodes)\\
                                           & \multirow{2}{*}{96 GB (3 nodes)}  & 870 GB (1 node)\\
\multirow{2}{*}{E5-2620 2.00GHz (6 nodes)} & \multirow{2}{*}{128 GB (8 nodes)}  & 1.61 TB (7 nodes)\\
                                           &                                   & 3.22 TB (4 nodes) \\ \hline

\end{tabular}
\end{table}

\section{Baseline: TeraSort for Suffix Array Construction}
\label{sec:terasort}
%why use TeraSort and bioinformatics
Taking the ostensible advantages of sorting large volumes of data and even including the optimizations~\cite{meno2011x1},
we argue that \ac{sa} construction with \ac{mr}
can only alleviate the intrinsic problems of scalability and efficiency in quantity
but cannot ameliorate them in quality.
It is because there exists an essential trait inherent in \ac{sa} construction with \ac{mr} that we cannot 
obviate---{\bf keeping every suffix in place}.  Without involving the optimizations for \ac{sa} construction,
TeraSort presents a simple method for analyzing the impact of such trait on scalability and efficiency.  
On the other hand, paired-end sequencing and alignment~\cite{illumina2016x1}---a popular technology in next-generation sequencing 
in bioinformatics---is adopted as our target application for analysis.
%As shown in Figure~\ref{fig:paired_end_dna}, paired-end read means that a DNA fragment is read twice from one
Paired-end read means that a DNA fragment is read twice from one
and the opposite directions.
We prepare two input files for the paired-end sequencing and alignment of the grouper genome, where one input file
contains those reads which are generated in one direction and the other input file contains those reads which
are generated in reverse order.  Each input file is about 32 GB in size and each read is about 200 bp.
All the suffixes of one input file is generated first for TeraSort and they are about 3.4 TB in size, which
is consistent with the {\bf self-expansion} factor ($\frac{1+200}{2} \approx 100$).
To make our baseline convincing and a fair comparison,
except the settings specified in Section~\ref{sec:mr}, we apply default settings to \ac{mr} and distribute
the input data in proportion to the sizes the disk space.  Suppose there are six input files in the same size,
we would distribute one to the nodes with 825 GB or 870 GB, two to the nodes with 1.61 TB, and three (instead of four) 
to the nodes with 3.22 TB to avoid too many mappers running on them.  Furthermore, we choose 64 reducers at most for the reason that each 
physical node can afford the moderate amounts of mappers ad reducers without crashing the systems.

%\begin{figure}[htbp]
%\centering
%\includegraphics[scale=0.3]{fig/paired_end_dna.eps}
%\caption{Illustration of paired-end read.}
%\label{fig:paired_end_dna}
%\end{figure}

\begin{figure}[htbp]
\centering
\includegraphics[scale=0.35]{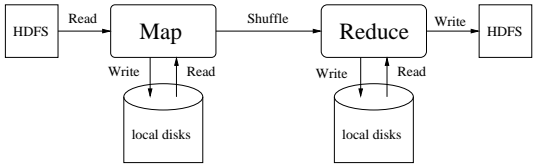}
\caption{The model for data store footprint in TeraSort.}
\label{fig:mr_suf}
\end{figure}

%model for MR handling suffix array construction
The execution time taken by a system, especially involving parallel computing, is subject to many factors intertwined 
with each other and hardly isolated completely.
So, the performance analysis with respect to time cannot
clearly reason about the requirements of the system for large-scale data processing.
Say, a system starts over a new task to finish the application after a task fails.
How do we decouple the effect of the failed task from the whole application running
when such case is non-deterministic?
Furthermore, what if the running application comprises several slow tasks?
More specifically, what is the exact execution time that this system is supposed to take?
In other words, we think execution time is still suitable for the judgement about the performance but seems unwieldy for
the performance analysis when large-scale data processing is considered.
Thus we decide to abandon the attempt to
evaluate how much time \ac{mr} takes to sort the suffixes.
Instead we propose an invariant and analytical abstraction commensurate with the
time that a system is supposed to take---tracking how much the effective data is
read from or written in the storages.
We call it {\bf data store footprint}, and the data constituting the output is
defined as the effective data.  In the previous case, the data associated with a failed task
doesn't count as the effective data if the output is not produced by it.
As such, data store footprint is deterministic and invariant.
To make the data store footprint analytical for \ac{mr}, we develop a model for data store footprint
shown in Figure~\ref{fig:mr_suf} based on the dataflow Map-Sort-Shuffle-Merge-Reduce.
The effective data in this model is categorized as the shuffling data, HDFS Read/Write, and Local Read/Write.

%observations
We are more focused on how Local Read/Write changes with the size of the input
since TeraSort doesn't change the sizes of the input, output and shuffling data.
Table~\ref{tbl:mr_terasort} shows the data store footprint of 5 different sized inputs.
In every one of these 5 cases, we regard {\bf the size of the input} (i.e. \ac{hdfs} Read) as {\bf 1 unit}
to analyze how many units TeraSort needs to sort the suffixes as the input size increases.
To be concise, we use {\bf XR} and {\bf YW} to represent {\bf X} units for {\bf reading}
and {\bf Y} units for {\bf writing} respectively,
where {\bf X} and {\bf Y} are arbitrary positive real numbers.
For example, Map in Case 1 performs 2.07W for local writing and the actual
size of the written data is $2.07\times 637.18 \approx 1318.96$ GB.  Similarly,
Reduce in Case 2 performs 1.37R for local reading and the actual
size of the read data is $1.39\times 1.24 \approx 1.72$ TB.
In addition, for the first 4 cases, we repeat each of them five times and produce two statistics, \ac{mean} and \ac{std}
to depict the $scalability_1$ of TeraSort and use Case 5 to point out its breakdown.
We make two observations about the data store footprint:

\begin{itemize}
  \item On Map-side, the load of Local Write is twice as much as the load of Local Read.
  \item The loads of Local Read and Write on Reduce-side are equivalent and increase as the input size increases.
\end{itemize}

\begin{table*}[htbp]
\cprotect\caption{Data store footprint of 5 different sized inputs for TeraSort with 32 reducers.  The elapsed time {\bf excludes the generation of suffixes}.
Note that four experiments of Case 5 {\bf don't complete} \ac{sa} construction due to the failures of some reducers, which in turn takes longer time
than one succeeded experiment.  We use all of them as the metrics.}
\label{tbl:mr_terasort}
\footnotesize
\center
\begin{tabular}{|l|c|c|c|c|c|c|c|c|c|c|}\hline
\multirow{2}{*}{Input size} & \multicolumn{2}{|c|}{Case 1} &\multicolumn{2}{|c|}{Case 2} & \multicolumn{2}{|c|}{Case 3} &\multicolumn{2}{|c|}{Case 4} &\multicolumn{2}{|c|}{Case 5*} \\\cline{2-11}
 & \multicolumn{2}{|c|}{637.18 GB} &\multicolumn{2}{|c|}{1.24 TB} & \multicolumn{2}{|c|}{1.86 TB} &\multicolumn{2}{|c|}{2.49 TB} &\multicolumn{2}{|c|}{3.37 TB}\\\hline
 & Map & Reduce & Map & Reduce & Map & Reduce & Map & Reduce & Map & Reduce  \\\cline{2-11}
Local Read      & 1.03 & 1.03 & 1.03  & 1.39 & 1.03 & 1.66 & 1.03 & 1.76 & 1.03 & 1.88 \\
Local Write     & 2.07 & 1.03 & 2.07  & 1.39 & 2.07 & 1.66 & 2.07 & 1.76 & 2.07 & 1.88  \\
\ac{hdfs} Read  & 1.00 &      & 1.00  &      & 1.00 &      & 1.00 &      & 1.00 &        \\
\ac{hdfs} Write &      & 1.01 &       & 1.01 &      & 1.01 &      & 1.01 &      & 1.01  \\\hline
Shuffle         & \multicolumn{2}{|c|}{1.03} & \multicolumn{2}{|c|}{1.03} & \multicolumn{2}{|c|}{1.03} & \multicolumn{2}{|c|}{1.03} & \multicolumn{2}{|c|}{1.03}  \\\hline
\multicolumn{1}{|l|}{Time (min.)} & \multicolumn{2}{|c|}{\ac{mean}=61.8; \ac{std}=1.30} & \multicolumn{2}{|c|}{\ac{mean}=143.4; \ac{std}=4.83} & \multicolumn{2}{|c|}{\ac{mean}=230.4; \ac{std}=12.30} & \multicolumn{2}{|c|}{\ac{mean}=312.0; \ac{std}=12.65} & \multicolumn{2}{|c|}{\ac{mean}=709.4; \ac{std}=95.55} \\\hline

\end{tabular}
\end{table*}

\begin{figure}[htbp]
\centering
\includegraphics[scale=0.5]{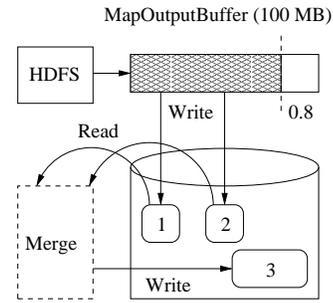}
\caption{Local I/O loading of TeraSort on the Map-side.}
\label{fig:mapper_rw}
\end{figure}

%analysis of the first observation
We delve deeply into the data store footprint by mining the counters of the mappers
and reducers to reason about what causes such amount of data read from and written in the local disks.
As shown in Figure~\ref{fig:mapper_rw}, the data in the buffer is spilled in the local disk once
the buffer reaches the level of 80\% (i.e. 80 MB).  Because the size of the input spilt for every
mapper is around 128 MB, every mapper spills the data to the local disk twice and two
intermediate files are merged into one later on.  For the sake of fault tolerance, the resulting
file resides in the local disk once a reducer fails to complete and an initiated reducer can
fetch the data.  Thus, there are approximate 1R and 2W for local disks on the Map-side.

\begin{figure}[htbp]
\centering
\includegraphics[scale=0.5]{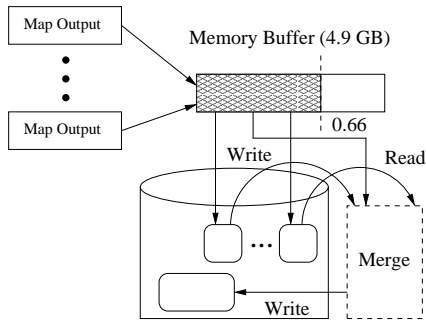}
\caption{Local I/O loading of TeraSort on the Reduce-side.}
\label{fig:reducer_rw}
\end{figure}

%analysis of the second observation
In Figure~\ref{fig:reducer_rw}, the memory buffer is determined by 70\% of the heapsize ($0.7 \times 7 = 4.9$ GB) and
a memory merger is triggered once the memory buffer reaches 66\% full.
In Case 1, a reducer receives the data of 20.56 GB and the number of spilled files is reckoned to
be around 6 ($20.56 \div 3.27 \approx 6$).  Since the default value of \verb#io.sort.factor# is 10,
all files residing in the local disk are sent to the \verb#Reduce()# without merging.  That is why there are
about 1R and 1W on the Reduce-side.  Due to the subtle mechanism of merging,
we estimate how much local disk I/O would be in Case 5 by the following steps:

\begin{enumerate}
  \item There are around 35 ($111.38 \div 3.27 \approx 34.06$) spilled files, where a reducer receives the data of 111.38 GB.
  \item In the first round, we merge 28 spilled files into 3 groups so there are 3 merged files and 7 spilled files left.
        Thus, $\frac{28}{34.06}$R/W is needed in this round.
  \item In the second round, we merge all the files into one file. 1R/W is needed in this round.
  \item The total units are $(\frac{28}{34.06} + 1)\times 1.03 \approx 1.88$.
\end{enumerate}

\begin{figure}[htbp]
  \begin{floatrow}
    \ffigbox{
      \includegraphics[scale=0.22]{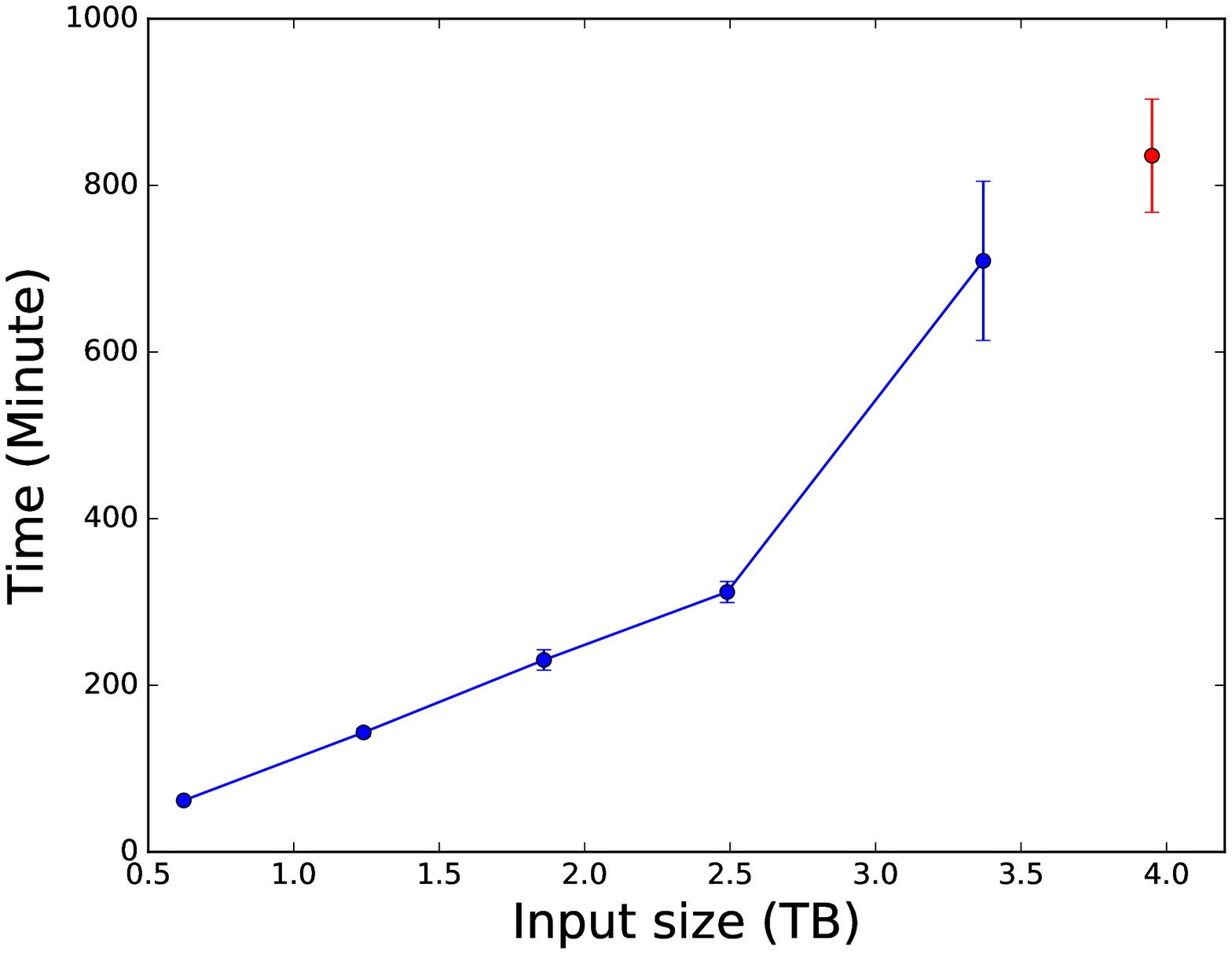}
    }{%
      \caption{$Scalability_1$ of TeraSort on examination. 
               The blue line depicts the trend with the \ac{mean} and \ac{std} in Table~\ref{tbl:mr_terasort}.
               The red point indicates the \ac{mean} and \ac{std} in Table~\ref{tbl:less_disk_terasort}.}
      \label{fig:scalability_1_terasort}
      \vspace*{-9pt}
    }
    \capbtabbox{
      \scriptsize
      \begin{tabular}{|c|c|c|}\hline
        \multirow{2}{*}{Input size}  &  \multicolumn{2}{|c|}{3.95 TB} \\\cline{2-3}
                                     & Map   & Reduce                 \\\hline
        Local Read                   & 1.03  & 1.85                   \\
        Local Write                  & 2.07  & 1.85                   \\
        \ac{hdfs} Read               & 1.00  &                        \\
        \ac{hdfs} Write              &       & 1.01                   \\\hline
        Shuffle                      & \multicolumn{2}{|c|}{1.03}                   \\\hline
        \multirow{2}{*}{Time (min.)} & \multicolumn{2}{|c|}{\ac{mean}=835.6}  \\
                                     & \multicolumn{2}{|c|}{\ac{std}=67.95}  \\\hline
      \end{tabular}
     }{%
       \caption{320 GB memory is managed by Yarn. 
                10 GB memory is allocated to every reducer of which the heapsize is 9GB.}%
       \label{tbl:less_disk_terasort}
     }
  \end{floatrow}
\end{figure}

%scalability 1
Figure~\ref{fig:scalability_1_terasort} illustrates the elapsed time of those 5 cases in Table~\ref{tbl:mr_terasort}
to examine the $scalability_1$ of TeraSort.  It shows that TeraSort scales in a linear sense from Case 1 to Case 4 but
no longer holds Case 5 in such linearity.
The failure of Case 5 is mainly caused by the errors about the memory issues such as \verb#GC overhead limit# or \verb#Java heap space#,
whereas we find that the increasing of the local disk I/O is endangering $scalability_1$ as well.  
As to the memory issues, TeraSort picks the first 10 bytes as the key to group the suffixes for sorting.
However, it is very common that plenty of suffixes are grouped together for sorting because their first 10 characters
are the same (e.g. \verb#ATATATATAT#), thereby stressing the heap space and \ac{gc} out.
In contrast, the lack of the enough disk space would compel TeraSort to start those reducers running on the nodes with
less disk space over on the other nodes, thereby taking more time to complete \ac{sa} construction.
Suppose two reducers in Case 5 are running at the same pace on the same node, their temporary files and outputs would
occupy about 644 GB ($111.38 \times 2 \times 2.89$) disk space which in turn is very likely to make the node unusable
and reschedule the reducers to the other available nodes.  Worse still, such the deficiency may cause non-deterministic
elapsed time.  In our environment, the resources of all the nodes are used only for our experiments to defer such the breakdown.

As illustrated in Table~\ref{tbl:less_disk_terasort}, to reinforce the hypothesis above, 
we increase both the memory and input size to eliminate the memory issues
and make the lack of the disk space more stressful for our cluster respectively.
Although the ratios of Local Read/Write is smaller than those of Case 5 due to the larger heapsize,
the size of the files generated by a reducer would be about 738 GB ($129.02 \times 2 \times 2.86$).
We do find that all failed reducers are caused by the lack of the enough disk space, which in turn affects
the completion time dramatically---the \ac{std} of the case in~Table~\ref{tbl:less_disk_terasort} is much
higher than those of Case 1 to 4.  Besides the non-deterministic completion time, we conclude that
on-disk merging is inevitable and would result in more extra local disk I/O on the Reduce-side
when the input size becomes larger.  Generally, as shown in Figure~\ref{fig:reducer_rw}, 
this is because there would be many files spilled from the memory buffer and then, merged into one single file
through more than one iteration if needed.

%\begin{figure}[htbp]
%  \begin{center}
%    {\subfigcapskip=1pt
%    \subtable[Feasibility of the manipulations concerning four conditions.]{
%      \label{tbl:four_conditions}
%      \scriptsize
%      \begin{tabular}{|c|c|c|}\hline
%        \multirow{2}{*}{Input size}  &  \multicolumn{2}{|c|}{3.95 TB} \\\cline{2-3}
%                                     & Map   & Reduce                 \\\hline
%        Local Read                   & 1.03  & 1.85                   \\
%        Local Write                  & 2.06  & 1.85                   \\
%        \ac{hdfs} Read               & 1.00  &                        \\
%        \ac{hdfs} Write              &       & 1.01                   \\
%        Shuffle                      &       & 1.03                   \\\hline
%        Time (min.)                  & \multicolumn{2}{|c|}{\ac{mean}=727; \ac{std}=10.14}  \\\hline
%      \end{tabular}}
%    }
%    \subfigure[Elpased time of Table~\ref{tbl:mr_terasort}]{\label{fig:scalability_1_terasort}\includegraphics[scale=0.15]{fig/scalability_1_terasort.eps}}
%    \vspace*{0pt}
%   %\hspace{5em}
%   %\subfigure[Performance Chart of Simulation]{\label{fig:performance}\includegraphics[scale=0.5]{fig/gnuplot_performance.eps}} \\
%  \end{center}
%  \caption{$Scalability_1$ of TeraSort on examination.}
%\end{figure}

%\begin{figure}[htbp]
%\centering
%\includegraphics[scale=0.3]{fig/scalability_1_terasort.eps}
%\caption{$Scalability_1$ of TeraSort on examination.}
%\label{fig:scalability_1_terasort}
%\end{figure}

%no performance tuning
Note that there is no performance optimization by parameter tuning for TeraSort.
Based on the analytical discussion above, there exits an essential trait
inherent in TeraSort that we cannot obviate---{\bf keeping every suffix in place}.
Under the influence of this trait, we argue that, for \ac{mr}, the innate capability
to handle the block devices and the superior mechanism of
message passing restrain TeraSort from scaling well, no matter how we tune the parameters for TeraSort.
This is because the necessity for keeping every suffix in place easily makes the
space requirement grow to the extent that the heavy I/O loads of the
local disks and the strong demand for the space to store those processing suffixes
degrade the performance severely.
We believe that any parameter tuning for the performance optimization without considering this issue would
not give the desired promise of the scalable \ac{sa} construction.

%However, how do we alleviate the I/O loads of local disks and abate the demand for the space?
% tributed systems
%Does there exists any better schemeto the : d abate the demand for the space
%to make \ac{sa} construction scale better than TeraSort.

\section{Scheme for Scalable and Efficient Suffix Array Construction}
%general charachteristics of suffix array construction in distributed system
To resolve the issue of keeping every suffix in place, we set a goal of
{\bf keeping only the raw data in place}, which means a suffix is
obtained via the query about it.
We examine the data store footprint of this goal specifically through three criteria:
data movement, data access pattern, and storage I/O.
In our goal, the data movement of suffixes would be more fine-grained since
the suffixes are generated, stored, and queried on the fly in tiny size. Besides,
random and irregularly frequent access to the raw data not only destroys the locality,
thereby making caching difficult, but also exacerbates the overhead of the storage I/O
and the network communication.  From the examination above, we find out that \ac{sa} construction is possessed
of the extreme scale of processing the data---generating each suffix (e.g. a few bytes) and
sorting all the suffixes (e.g. several Terabytes).
It indicates that relying on only one type of storage (e.g. disk) might improve the performance of one extreme
by sacrificing the performance of the other extreme.

%memory/disk storage
%why not in-memory file system
Since \ac{mr} has proven its great capability of sorting, we conceive of an abstraction
that requires the space for only the raw data and is competent to access the suffixes at speed
by taking advantage of memory.  Having the benefits of memory, we can overcome the
problem of the data access pattern and alleviate the I/O loads by trading disk I/O for memory I/O.
Though the distributed in-memory file systems, Alluxio (formerly known as Tachyon) for example, are popular
for accelerating large-scale data processing by exploiting the better speed of data access
in memory.  However, the structure behind the distributed in-memory file systems still emphasizes
the management of files in a way that considers the underlying storages as the block devices.
The drawback to the block devices is that, given an index of a suffix, it takes time to seek the target block, retrieve
the whole block, get the wanted suffix, and discard the other data in that block.
Moreover, we can infer from the extreme scale we just describe above that 
the distributed in-memory file systems can indeed help to enhance the performance of \ac{sa} construction but
the majority of access time would be made redundant which results in poor scalability.

\begin{figure*}[htbp]
  \begin{center}
    \subfigure[The model for data store footprint in our scheme.]{\label{fig:mr_in_memory_kv}\includegraphics[scale=0.35]{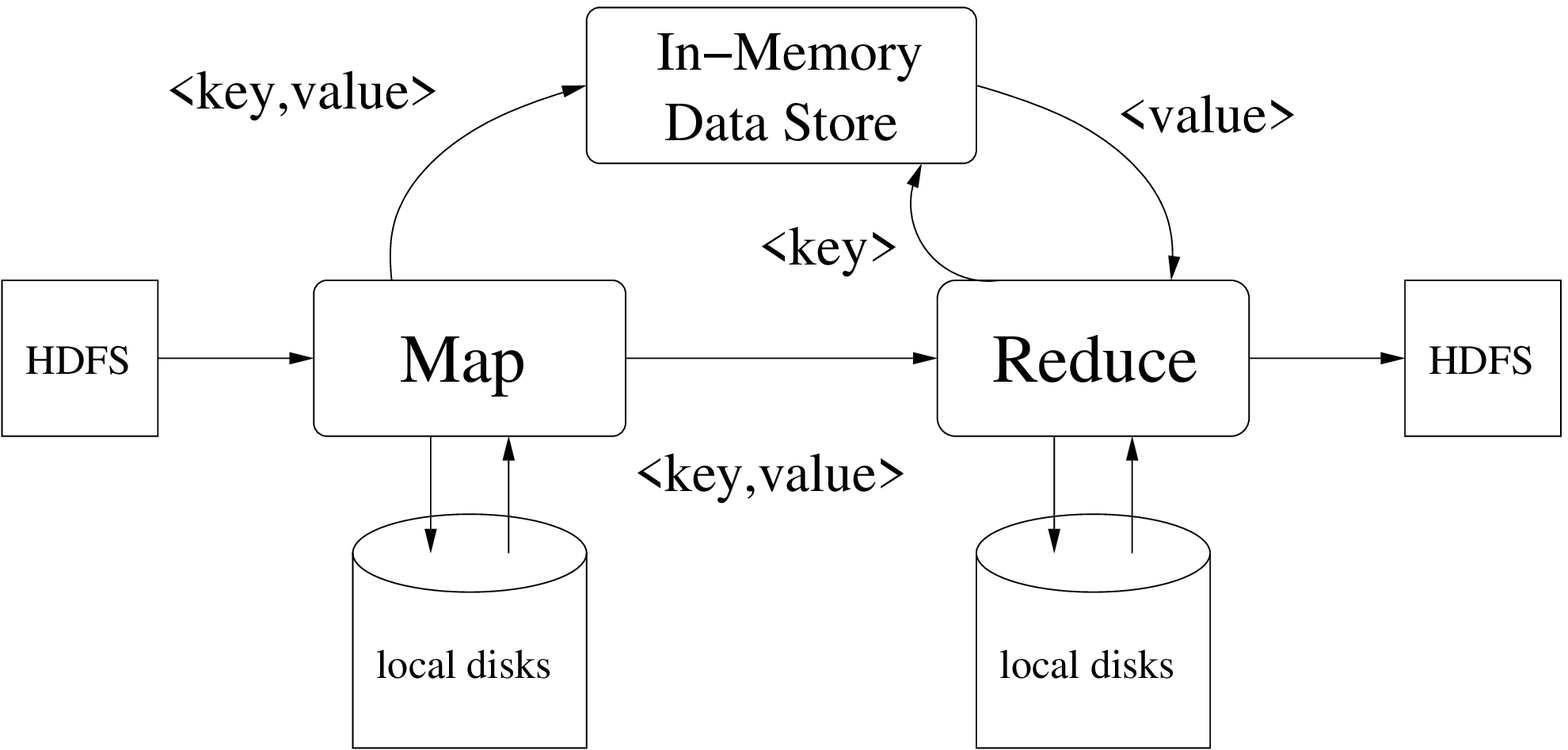}}
    \hspace{2em}
    \subfigure[Illustration of how {\bf keeping only the raw data in place} is realized.]{\label{fig:in_mem_ds}\includegraphics[scale=0.27]{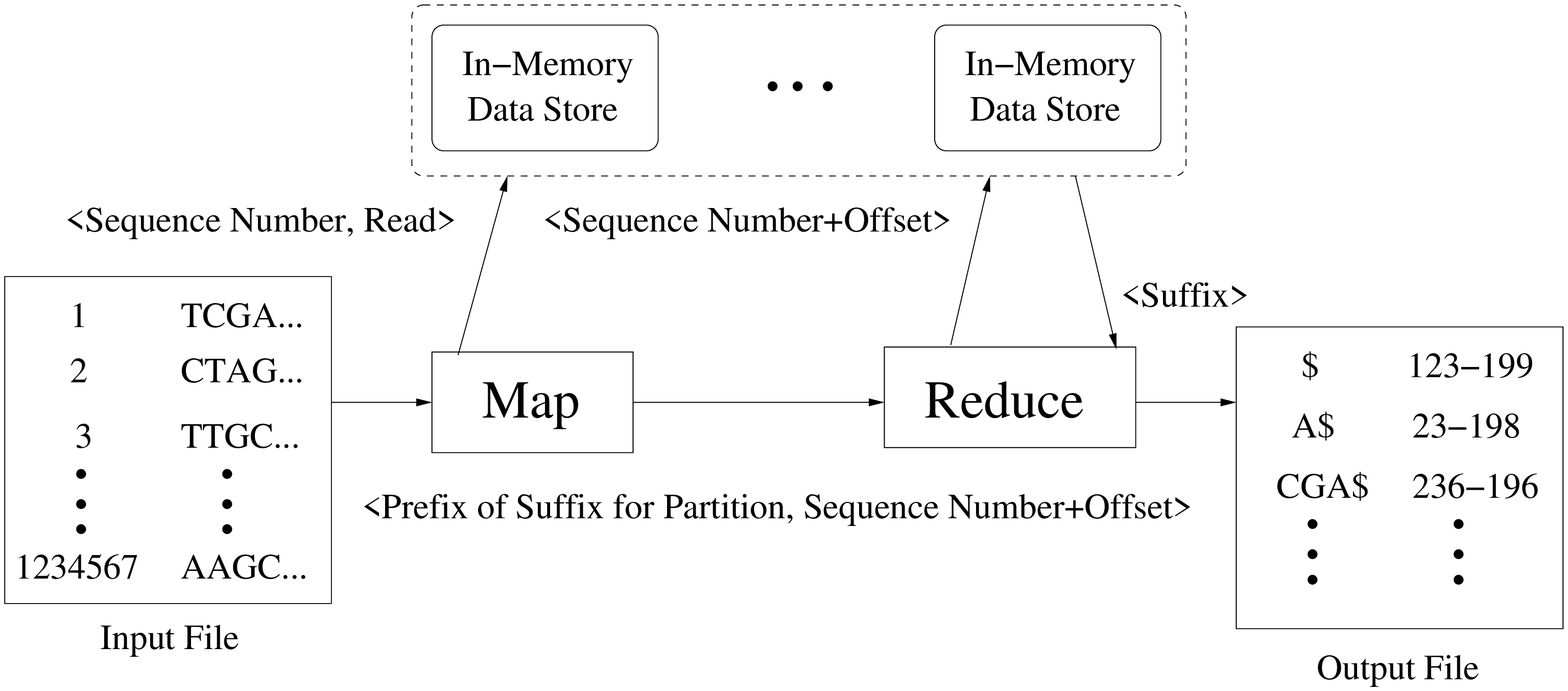}}
   %\subfigure[Performance Chart of Simulation]{\label{fig:performance}\includegraphics[scale=0.5]{fig/gnuplot_performance.eps}} \\
  \end{center}
  \vspace*{-18pt}
  \caption{Scheme for Scalable and Efficient Construction of Suffix Array.}
\end{figure*}

%in-memory key-value data store
Instead of the distributed in-memory file system, we adopt the distributed in-memory data store system like
Redis~\cite{redis} as the realization of our abstraction.
As an in-memory key-value data store system for small chunks of arbitrary data, Redis is natural to be
integrated into \ac{mr} through the communication of key-value pairs and easy to scale in number and size.
Figure~\ref{fig:mr_in_memory_kv} illustrates the data store footprint in our scheme.
In the following sections, we introduce our scheme in detail and assess its possibilities from a standpoint of 
%$Scalability_1$ and $Scalability_2$.
{\bf load scalability}, {\bf structure scalability}, {\bf space scalability} and {\bf efficiency}.

\subsection{Load Scalability}

% cooperation of storage, load balance
If a system is possessed of load scalability, it can function gracefully at
light, moderate, or heavy loads while making good use of available resources.
We rephrase it as balancing the loads on the memories, disks, and network communication
by the cooperation among them.  In other words, our scheme offloads keeping every suffix in place with disks
onto the concept of keeping only the raw data in place with memories and network communication to achieve 
better load scalability.  Figure~\ref{fig:in_mem_ds} conveys this concept that the in-memory
data store system (i.e. a bunch of Redis instances) takes charge of storing the raw data in memories and responding the queries about suffixes 
via network communication,
whereas \ac{mr} only needs to manage the indexes of the suffixes to abate the loading of local disk I/O since
the size of the indexes is relatively smaller than the size of suffixes.  
The same format as what we use for TeraSort, the first and second columns in Input File in Figure~\ref{fig:in_mem_ds} 
are full of the sequence numbers and reads respectively.
To distribute the reads to the Redis instances evenly, we make every sequence number modulo the number of the Redis instances
to determine which Redis instance the key-value pair \verb#<Sequence Number, Read># goes to.
In addition to putting the raw data in the Redis instances,
Map sends the indexes of the corresponding suffixes to Redcue so that Reduce can 
acquire them from the Redis instances using the indexes.
To make the number of suffixes be dispatched to each reducer evenly,
our scheme adopts the partitioning method similar to TeraSort and~\cite{meno2011x1} by assuming the randomness of the suffixes.
Given the number of reducers (e.g. $n$), we sample
$N$ suffixes and sort them to estimate the ranges of the suffixes, where $N$ is $10000 \times n$.
In our case, after sampling 320000 suffixes and sorting them,
we pick the 10000th, 20000th, ..., and 310000th suffixes to determine the boundaries of the ranges.
Finer partition can be achieved by increasing the number of sampling points.
Note that our scheme overcomes the self-expansion by shifting it from disks to memories and network communication.
Through the speed of memory access and network communication, our scheme can enhance the load scalability which
outweighs the decrease of available memories.  We prove it in Section~\ref{sec:analysis}.

%\begin{enumerate}
%\begin{itemize}
%  \item It doesn't impair the load scalability.
%  \item It is not sensitive to the growth of the input size.
%\end{itemize}
%\end{enumerate}

\subsection{Structural Scalability}
If a system is possessed of structural scalability, its implementation or standards
do not impede the growth of the input size.  We investigate structural scalability
in our scheme with the following requirement: it is relatively insensitive to the growth of the input size
with respect to TeraSort.  By means of these Redis instances, our scheme encapsulates
those suffixes in the raw data and handled them on demand to reduce the self-expansion effect on
the suffixes in exchange for the self-expansion effect on the indexes of those suffixes.
Here comes the question: wouldn't such self-expansion destroy the scalability like what just happened to TeraSort?

\begin{figure}[htbp]
\centering
\includegraphics[scale=0.5]{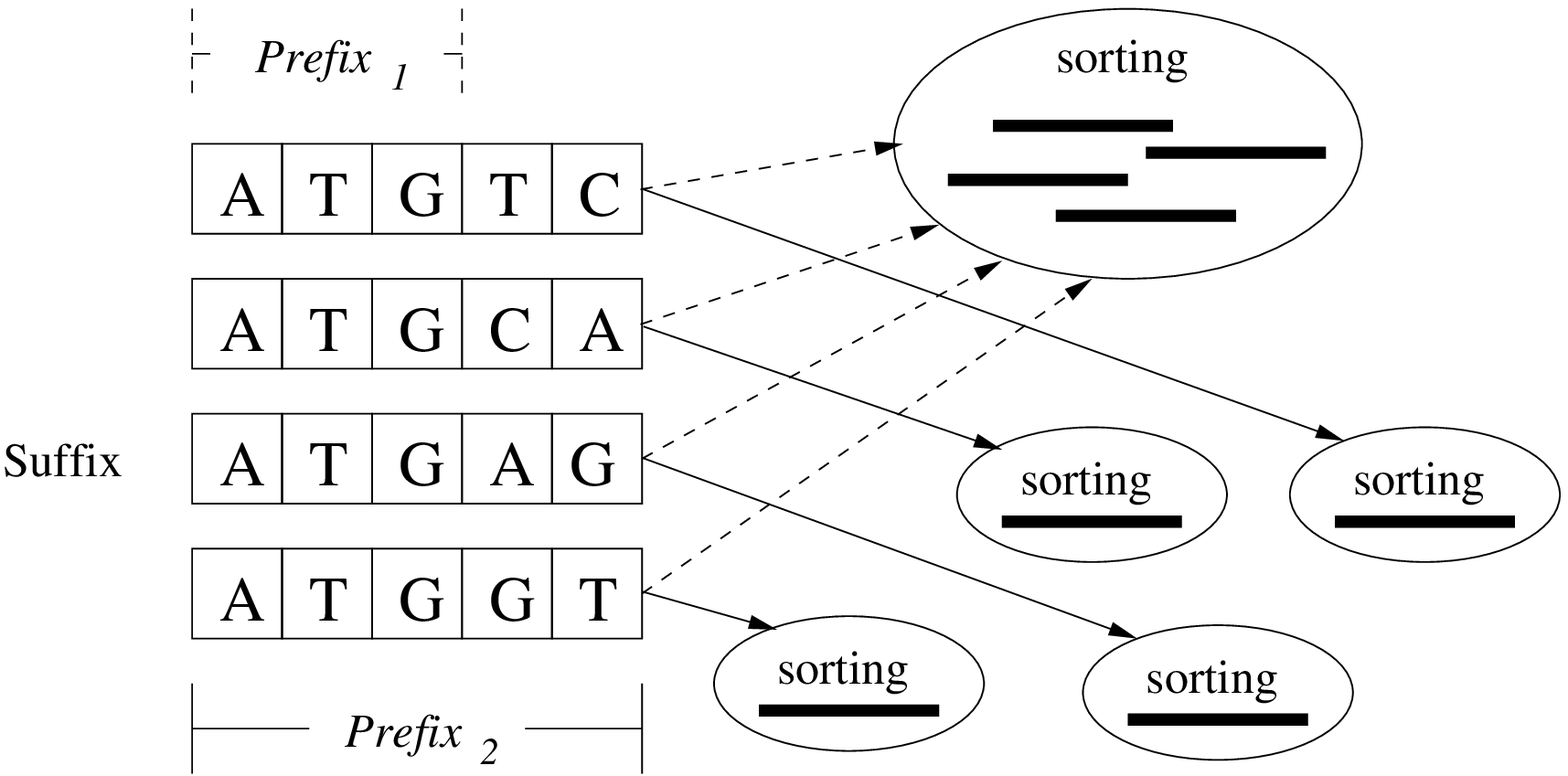}
\caption{How different length of the prefixes determines the number of the groups and the size of the group for sorting.}
\label{fig:prefix_group}
\end{figure}

%nmerical representation for key
The index we mean here is the key-value pair communicating in \ac{mr} and can be used to acquire the corresponding
suffix.  As shown in Figure~\ref{fig:in_mem_ds}, once we know the \verb#Sequence Number#, we look it up
in the target Redis instance, find the read, and extract the suffix from the read by the \verb#offset#.
Rather that using \verb#String#, we choose the numerical representation in \verb#long# or
\verb#int# due to its better capability of accommodating more objects to address within a fixed number of bytes.
In addition, it is also flexible to expand as the input size increases with very little overhead (e.g. a few bytes).
The exact way to represent \verb#Sequence Number# and \verb#offset# numerically is \verb#Sequence Number#
$\times 1000 +$ \verb#offset# since \verb#offset# ranges from 0 to 200.  The retrieval of \verb#Sequence Number#
and \verb#offset# can be done by division and modulo respectively.
On the other hand, there are only five possible characters in a read: A, C, G, T and \$.  With base 5, we
use \$=0, A=1, C=2, G=3 and T=4 to represent the key used in \ac{mr} numerically.  To fit \verb#long# or
\verb#int#, we encode the prefix of every suffix in a fixed number of characters (e.g. 10).
So do the boundaries of the ranges for the partition.  Moreover,
there is every chance that a lot of the short suffixes are grouped together for sorting,
which more likely results in the errors about \verb#GC overhead limit# or \verb#Java heap space#.
We discover that if the length of a suffix is smaller than the length of the prefix we defined,
the prefix is actually the suffix itself.
Say, the prefix in a length of 10 characters for a suffix \verb#AGT$# is \verb#AGT$# itself.
Reaping such the benefit, our scheme doesn't have to sort those suffixes because they are the same,
which provides the memory relief for the reducers and saves time.
Figure~\ref{fig:prefix_group} illustrates how our scheme partitions the sorting group and how the size
of a sorting group changes according to the length of the prefix.  Since $Prefix_1$ is in a length of 3,
the prefixes of those four suffixes are all \verb#ATG#.  Conforming to \ac{mr}, these suffixes are grouped
together and sorting at the same time.  Applying $Prefix_2$, we would have four sorting groups and
each contains one suffix.  The order among these sorting groups is maintained by the partitioner, thereby
matching up with the order in the scenario of $Prefix_1$.
There is a rule of thumb: the longer length of the prefix, the smaller size of the sorting group which 
in turn requires less memory to sort the suffixes.

%network communication
Acquiring the suffixes one by one through the network communication squanders the time we gain
from the disk I/O and makes the memory I/O busy.  To utilize the network communication and memory I/O efficiently,
our scheme aggregates those indexes of the suffixes which are stored in the same Redis instance,
and retrieves the suffixes from it at one time to save the communication cost.
Since Redis doesn't support the retrieval of the multiple partial contents,
we add a new Redis command called ''\verb#mgetsuffix#''~\cite{mgetsuffix_redis} in Redis and its corresponding function in 
Jedis~\cite{mgetsuffix_jedis} to retrieve the whole suffixes back instead of the whole reads.
As such, our scheme almost saves half an amount of data communicating in the network while acquiring the suffixes.
Because putting multiple reads at one time is permitted in Redis, our scheme lets the mappers aggregate
those reads which are assigned to the same Redis instance and put them to it when the mappers finish
reading the input file. 

In summary, our scheme further relives the self-expansion effect on the suffixes by aggregating as many suffixes as possible
while storing the reads and acquiring the suffixes.  As such, not only is the access of the memory and network I/O reduced, 
but the utilization of the network bandwidth is improved.
Concerning the self-expansion effect on the indexes of the suffixes, our scheme has superior capability of restricting
it in a constant factor unless there exists some sorting group that cannot fit into the memory for sorting.
For example, \verb#int# contains four byte so the threshold is 13 because the numerical value of \verb#TTTTTTTTTTTT#
is 1220703124, which is the largest number smaller than 2147483647.  So the total bytes of a key-value pair used in \ac{mr}
is 12 bytes (i.e. \verb#int#$+$\verb#long#), which is smaller than 100 by a factor of 8 and has nothing to do with
the length of reads.  Only when some sorting group is too big to fit into the memory, do we need to partition the sorting
groups in finer grain by lengthening the prefix.  If we replace \verb#int# with \verb#long# to accommodate the longer prefix 
(the threshold would be 26), the total bytes are just 16 bytes, which is still smaller than 100 by a factor of 6.

\subsection{Space Scalability}

%the size of the sorting group
If a system is possessed of space scalability, its memory requirements do not grow to intolerable levels as
the number of items it supports increases.  Here it is referred to the economical usage of the heap and
two factors are considered: {\bf the size of the sorting group} and {\bf the type of the garbage collector}.
Involved in TeraSort as well, the former factor usually results in \verb#Java heap space#
while a reducer attempts to allocate the memory for the big sorting group, or \verb#GC overhead limit# 
when little progress is made on the \ac{gc}.
On the other hand, if the sizes of the sorting groups are very small, a reducer would waste time on the 
overhead of switching from group to group for sorting only the small number of the suffixes.  Furthermore, the amount
of suffixes acquired from the Redis instance would also be very small, which means low throughput.  This dilemma comes from
the fact that the size of a sorting group varies all the time, which in turn influences the sorting time and the throughput.
To make the sizes vary within a narrow range, we accumulate the sorting groups without sorting until such accretion exceeds
some threshold.  That is, we prevent not only the heap from the shortage of available memory by
shrinking the sorting group size but also the time of sorting from the switching overhead by collecting the sorting groups together.
We choose $1.6 \times 10^6$ as the threshold value just because the experiments with this value are better than $3.2 \times 10^6$
and $8 \times 10^5$.  In our scheme, the sorting would not be triggered until the number of suffixes is more than
the threshold value.

%Garbage Collection
It is inevitable to allocate and free memory over and over for \ac{sa} construction on a large scale.
Even though we can utilize the heap economically without any error, there exists a performance issue that
the throughput of the suffixes acquired from the Redis instance is decreased when {\em stop-the-world} \ac{gc}
is performed, which means the execution of the application is completely suspended during the \ac{gc}.
More specifically, it mainly occurs at the moment that we need the space for the new suffixes but there is less
space in the heap because the past suffixes occupy the heap.  Once {\em stop-the-world \ac{gc}} starts to clean them, 
it pauses for acquiring the suffixes, thereby deteriorating the throughput.
Omitting the details of the explanations,  our scheme chooses 1 GB for {\em young generation} and the rest of the heap
for {\em old generation} so that the past suffixes can be moved to {\em old generation} soon and we use
\verb#-XX:+AlwaysTenure# as a catalyst to make this procedure quicker.  With \verb#-XX:+UseConcMarkSweepGC#, 
our scheme can sweep the past suffixes off {\em old generation} massively and acquire the new suffixes concurrently, thereby
improving the throughput.  It is because the suffixes would not be reused after sorting that moving the used suffixes
to {\em old generation} as soon as possible could save time on triggering \ac{gc} and make more space available at one time.

\subsection{Analysis and Efficiency of Our Scheme}
\label{sec:analysis}
{\bf We let our scheme execute on the same \ac{mr} environment presented in Section~\ref{sec:terasort}}
for the reason that the enhancement of the scalability and performance can be clearly reflected in 
data store footprint and time respectively.  As such, we let every node donate the extra memory for its Redis instance
to accommodate the input files and set the length of the prefix 23.
The overhead of storing the input data in these 16 Redis instances is about 1.5 times as much space
as the input size due to the metadata.  For instance, those 16 Redis instances need 48 GB memory to
store the input data of 32 GB in size (i.e., each node has to donate the extra 4 GB memory for that).
In Table~\ref{tbl:mr_in_mem_ds}, {\bf we normalize the metrics by regarding the size of output as 1.01 unit} for
intuitively comparing the workload on the disk I/O.  This is because the outputs of TeraSort and our scheme must
be the same so we use the output size as the reference point.
Note that the input data of Case 1 to 5 is exactly the same as those in Table~\ref{tbl:mr_terasort}.

%how our scheme relieve local I/O loading on the Map-side, 
The key and value are all represented in \verb#long# so the size of one key-value pair is 16 bytes.
%On the other hand, the size of the metadata associated with one key-value pair is 16 bytes, which means its size matters now.
In one InputSplit (or one mapper), the number of the records is 639,893 in average but the number of key-value pairs that a mapper would emit is
about $639,893 \times 200$.  The total size of the key-value pairs would be 1.95 GB.
Similar to the induction in Figure~\ref{fig:mapper_rw}, there are around 50 spilled files that need $1 + \frac{45}{50}$ units for Local Read 
and $2 + \frac{45}{50}$ units for Local Write where 1 unit represents 1.95 GB.  The ratio $\frac{Local\ Read}{Local\ Write}$ is verifiable
in Table~\ref{tbl:mr_in_mem_ds}.  Since our scheme includes the generation of suffixes, it takes more time in Map (25 mins in average) 
than TeraSort does.
Nevertheless, we can see the significant reduction of the local disk I/O in contrast to TeraSort.
This is because \ac{mr} handles only the indexes of the suffixes rather than the suffixes.

\begin{table*}[htbp]
\cprotect\caption{Data store footprint of our scheme with 32 reducers and the elapsed time {\bf includes the generation of suffixes}.
Note that Case 6 is the \ac{sa} construction for the pair-end sequencing and alignment with two input files.}
\label{tbl:mr_in_mem_ds}
\footnotesize
\center
\begin{tabular}{|l|c|c|c|c|c|c|c|c|c|c|c|c|}\hline
\multirow{2}{*}{Input size} & \multicolumn{2}{|c|}{Case 1} &\multicolumn{2}{|c|}{Case 2} & \multicolumn{2}{|c|}{Case 3} &\multicolumn{2}{|c|}{Case 4} &\multicolumn{2}{|c|}{Case 5} &\multicolumn{2}{|c|}{Case 6} \\\cline{2-13}
 & \multicolumn{2}{|c|}{5.86 GB} &\multicolumn{2}{|c|}{11.72 GB} & \multicolumn{2}{|c|}{17.57 GB} &\multicolumn{2}{|c|}{23.43 GB} &\multicolumn{2}{|c|}{31.76 GB} &\multicolumn{2}{|c|}{63.12 GB} \\\hline
 & Map & Reduce & Map & Reduce & Map & Reduce & Map & Reduce & Map & Reduce & Map & Reduce \\\cline{2-13}
Local Read      & 0.30 & 0.16 & 0.30  & 0.16 & 0.30 & 0.16 & 0.30 & 0.16 & 0.30 & 0.16 & 0.30 & 0.16 \\
Local Write     & 0.45 & 0.16 & 0.45  & 0.16 & 0.45 & 0.16 & 0.45 & 0.16 & 0.45 & 0.16 & 0.45 & 0.16 \\
\ac{hdfs} Read  & 0.01 &      & 0.01  &      & 0.01 &      & 0.01 &      & 0.01 &      & 0.01 &     \\
\ac{hdfs} Write &      & 1.01 &       & 1.01 &      & 1.01 &      & 1.01 &      & 1.01 &      & 1.01  \\\hline
Shuffle         & \multicolumn{2}{|c|}{0.16} & \multicolumn{2}{|c|}{0.16} & \multicolumn{2}{|c|}{0.16} & \multicolumn{2}{|c|}{0.16} & \multicolumn{2}{|c|}{0.16}& \multicolumn{2}{|c|}{0.16} \\\hline
\multicolumn{1}{|l|}{Time (min.)} & \multicolumn{2}{|c|}{\ac{mean}=63.2; \ac{std}=0.45} & \multicolumn{2}{|c|}{\ac{mean}=100.0; \ac{std}=0.71} & \multicolumn{2}{|c|}{\ac{mean}=156.6; \ac{std}=2.41} & \multicolumn{2}{|c|}{\ac{mean}=203.4; \ac{std}=4.16} & \multicolumn{2}{|c|}{\ac{mean}=284.2; \ac{std}=8.38} & \multicolumn{2}{|c|}{\ac{mean}=647.0; \ac{std}=12.19} \\\hline

\end{tabular}
\end{table*}

%shuffling and network usage
Handling the indexes of the suffixes also helps to decrease the amount of shuffling data.
Some might argue that the network usage in our scheme is higher than TeraSort because there is
extra network communication of the suffixes generated from those Redis instances.  Actually, it is 
the procedure of generating the suffixes and necessary for TeraSort as well.  The reason why we exclude
the time of generating the suffixes from TeraSort is fairness---writing the suffixes into the disk takes a lot of time.
Once we take the generation of the suffixes into account, our scheme really reduces the network usage for shuffling
by comparison with TeraSort.

\begin{figure}[htbp]
\centering
\includegraphics[scale=0.47]{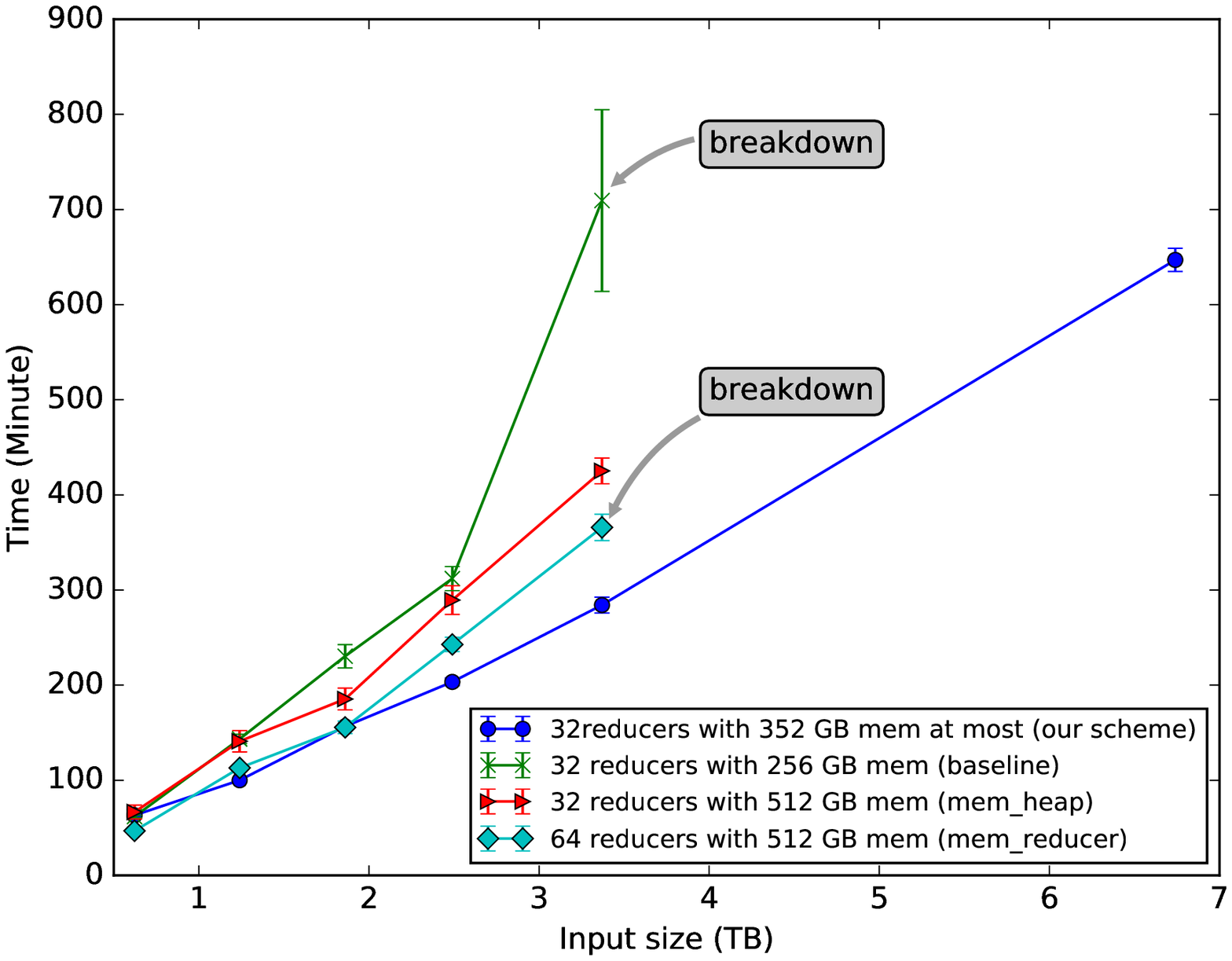}
\caption{$Scalability_{1,2}$ of TeraSort and our scheme on examination.}
\label{fig:scalability_our_scheme}
\end{figure}

%reduce-side
Benefiting from the smaller key-value pairs, the reducers don't need to merge the spilled files
more than one iteration.  Taking Case 5 as an example, one reducer receives about 17 GB
data ($\frac{3.4 TB\times 0.16}{32} \approx$ 17 GB) and there are only 6 spilled files ($\frac{17}{2.8} \approx 6$)
which we can merge in one iteration.  That's why the Local Read/Write of Reduce is the same as
the Shuffle.  The throughput of the suffixes acquired by a reducer is about 20 MB/sec and doesn't last for
the whole time.  We think the network bandwidth (1 Gigabit in our scheme) is not fully utilized and could be improved
by increasing the number of the reducers or the size of the sorting group.
We roughly classify the computation time into three categories---getting suffixes, sorting, and others---where
their percentages are about 60\%, 13\%, and 27\% respectively.  Since the latency of acquiring the suffixes
is the dominant factor, the high-end network like InfiniBand would be very helpful to our scheme.
Nevertheless, as shown in Figure~\ref{fig:scalability_our_scheme}, 
our scheme (blue) outperforms TeraSort (green) in terms of time and space as the input size increases regardless of
the inclusion of generating suffixes.
We believe that the access of the suffixes through the memory and network with the in-memory data store system is much better 
and more compact than through the block devices with the conventional filesystem.

%\subsection{Efficiency}
%\label{sec:efficiency}
%efficiency
While the better $scalability_1$ of our scheme is qualitatively proved and exhibited in quantity, some might argue
our scheme is supposed to be better because of the extra memory, and claim the $scalability_1$ of TeraSort can also be 
improved using the same way.  As discussed in Section~\ref{sec:terasort}, the memory issues and deficiency of disk space could be
appeased by including more memory in \ac{mr} but the question is---how efficient is the way you add the extra memory?
There are two straightforward ways to take advantages of the memory: 
increasing either the heap size of the reducers or the number of the reducers;
we call them $mem_{heap}$ and $mem_{reducer}$ respectively.
Regarding Table~\ref{tbl:mr_terasort} as the baseline,
we evaluate the efficiency of $mem_{heap}$, $mem_{reducer}$, and our scheme with respect to the amount of memory and speedup.
The total memory for $mem_{heap}$ and $mem_{reducer}$ is 512 GB. 
Table~\ref{tbl:mr_32G_32_reducers} and~\ref{tbl:mr_32G_64_reducers} show the experimental results of the same 5 cases in Table~\ref{tbl:mr_terasort}.

%scalability model
Given the input size $x$, we interpret the elapsed time as a simple model $f(x)$ below.
This model consists of the linear part $ax+b$ and non-deterministic part $N/A$ where $a$, $b$ and $breakdown$ are the parameters.
Here $breakdown$ is the threshold value of the input size that $scalability_1$ falls into ruin.
The parameter $a$ and $breakdown$ are associated with $Scalability_1$, whereas the parameter $b$ that lumps the parallelization 
and acceleration together, for example the number of reducers and the frequency of a CPU respectively, is associated with $scalability_2$.
It seems that $b$ is supposed to depend on $x$ more or less but we regard it as a constant for simplicity to get more insights.

\begin{equation*}
\label{equ:cross_entropy}
\small
  f(x) =
  \begin{cases}
    ax+b       & \quad \text{if } x < \text{$breakdown$}\\
    N/A        & \quad \text{otherwise}\\
  \end{cases}
\end{equation*}

%mem_heap
In Table~\ref{tbl:mr_32G_32_reducers}, the local disk I/O of Case 1 and 2 is exactly the same as the one in Table~\ref{tbl:mr_terasort} which means
both $mem_{heap}$ and the baseline are good at ease with their input files of Case 1 and 2.  As the input size increases, 
$mem_{heap}$ undoubtedly maintains better $sclability_1$ than the baseline, even having the bonus for the moderate input size like Case 3 and 4.
Interestingly, if we extrapolate the green line from Case 4, we find the red line presumably approaches the green line and 
continues the slope of the green line.
In other words, $mem_{heap}$ defers the $breakdown$ of the baseline's $scalability_1$ but doesn't change its $a$ at last.
Although less local disk I/O means less time, the bonus, attributed to the acceleration done by the larger heap, vanishes at
the time when the heap starts to suffer the deficiency of the memory.  

With the parallelization by doubling the number of the reducers, $mem_{reduce}$ displays the promising metrics
until the breakdown shown in Table~\ref{tbl:mr_32G_64_reducers}.  By dispensing the data to the more reducers,
such the parallelization makes every reducer process the half size of data to decrease the local disk I/O and the completion time.
As we can see in Figure~\ref{fig:scalability_our_scheme}, the slope of the cyan line is similar to the slope of the green line.
Despite the fact that the parallelization reduces the local disk I/O, the $breakdown$ is exactly the same as the $breakdown$ in the baseline.
It is because the parallelization couldn't alter the size of the sorting groups that the $scalability_1$ of $mem_{reduce}$ still 
suffers the fragility.  The effect of the parallelization is reflected in $b$ as what we expect.

\begin{table}[htbp]
\cprotect\caption{Data store footprint of $mem_{heap}$ with 32 reducers of which each uses 16 GB physical memory and 15 GB heap.
We show only Local Read/Write on the Reduce-side because the other parts are the same as Table~\ref{tbl:mr_terasort}.}
\label{tbl:mr_32G_32_reducers}
\footnotesize
\center
\begin{tabular}{|l|c|c|c|c|c|}\hline
\multirow{2}{*}{Input size} & Case 1    &Case 2  & Case 3  &Case 4  &Case 5 \\\cline{2-6}
                            & 637.18 GB &1.24 TB & 1.86 TB &2.49 TB &3.37 TB \\\hline
Local Read                  & 1.03      &1.03    & 1.02    &1.33    &1.53\\
Local Write                 & 1.03      &1.03    & 1.02    &1.33    &1.53\\\hline
\multirow{2}{*}{Time (min.)} & \ac{mean}=66.6& \ac{mean}=141 & \ac{mean}=185.4& \ac{mean}=289.4& \ac{mean}=425.2 \\
                            & \ac{std}=7.30 & \ac{std}=11.22 & \ac{std}=11.48 & \ac{std}=15.04 & \ac{std}=13.55 \\\hline

\end{tabular}
\end{table}

\begin{table}[htbp]
\cprotect\caption{Data store footprint of $mem_{reducer}$ with 64 reducers of which each uses 8 GB physical memory and 7 GB heap.
{\bf The breakdown occurs in Case 5}: two experiments succeed in \ac{sa} construction, whereas
the other three fail due to the oversize sorting group.}
\label{tbl:mr_32G_64_reducers}
\footnotesize
\center
\begin{tabular}{|l|c|c|c|c|c|}\hline
\multirow{2}{*}{Input size} & Case 1    &Case 2  & Case 3  &Case 4  &Case 5* \\\cline{2-6}
                            & 637.18 GB &1.24 TB & 1.86 TB &2.49 TB &3.37 TB \\\hline
Local Read                  & 1.03      &1.03    & 1.03    &1.38    &1.56\\
Local Write                 & 1.03      &1.03    & 1.03    &1.38    &1.56\\\hline
\multirow{2}{*}{Time (min.)} & \ac{mean}=46.8& \ac{mean}=100 & \ac{mean}=156.6& \ac{mean}=242.8& \ac{mean}=365.8 \\
                            & \ac{std}=3.56 & \ac{std}=0.7 & \ac{std}=2.41 & \ac{std}=7.53 & \ac{std}=13.83\\\hline

\end{tabular}
\end{table}

%efficiency
We claim that the $scalability_{1,2}$ of our scheme is radically different from those of TeraSort.
Compared with $mem_{reducer}$, our scheme apparently presents smaller value of $a$, almost the same value of $b$,
and---the most vital---bigger value of $breakdown$.  On the other hand, with such $a$ and $b$, the speedup of our scheme 
is getting more significant while the input size is getting larger.
Furthermore, our scheme is more efficient than TeraSort with respect
to the memory usage.  In table~\ref{tbl:efficiency}, we list the efficiency of Case 1 to 4 but ignore Case 5 due to its
unstable metrics, where the \ac{mean} is adopted in the calculation of $speedup$.  Neither $mem_{heap}$ nor $mem_{reducer}$
can achieve more than 75\% efficiency.  In contrast, our scheme performs amazing efficiency, even greater than 100\%.
This is because the memory used for storing the input data is relatively small whereby the $mem_{ratio}$ is close to 1.
Although it is strange that the efficiency could be greater than 100\%, it is another evidence that the scalability of
our scheme is essentially different from that of TeraSort.  Say, if our scheme with 32 reducers is considered as the baseline, 
the efficiency of our scheme with 64 reducers would not be greater than 100\% since their scalability is structurally the same.

\begin{table}[htbp]
\cprotect\caption{The efficiency is calculated by $\frac{speedup}{mem_{ratio}}$.
The $mem_{ratio}$ of our scheme varies as the input size changes. }

\label{tbl:efficiency}
\small
\center
\begin{tabular}{|l|c|c|c|c|}\hline
                     & Case 1    &Case 2  & Case 3  &Case 4  \\\Xhline{1.2pt}
{\bf $mem_{heap}$}         & 46.4\%    &50.9\%   & 62.1\%   &53.9\%   \\
$mem_{reducer}$      & 66.0\%    &63.5\%  & 74.0\%  &64.3\%    \\\hline
$mem_{ratio}$        &\multicolumn{4}{|c|}{$\frac{512}{256} = 2$}   \\\Xhline{1.2pt}
Our scheme           & 95.5\%    &140.0\%   & 141.1\%  &134.5\% \\\hline
$mem_{ratio}$        & $\frac{256+9}{256}$ & $\frac{256+18}{256}$  & $\frac{256+27}{256}$    & $\frac{256+36}{256}$   \\\hline

\end{tabular}
\end{table}

It is worth noting that our scheme could be faster by not writing the suffixes into \ac{hdfs}.
This is because the suffixes can be obtained through the Redis instances with their indexes.
In other words, our scheme can also save the time in the following stages by exploiting 
the concept of keeping only the raw data in place.  The reason why we write them out is
for the fair comparison with TeraSort.

%Same as TeraSort, we don't optimize the performance of our scheme either.
%From an angle of the qualitative change in scalability, our scheme is a structural optimization that incorporates
%the additional memory and Redis into \ac{mr} so that the scalaility for \ac{sa} construction is enhanced through
%the reduction of local I/O in quantity.

\section{Conclusion}
We have presented a quantitative analysis of \ac{sa} construction in detail,
and qualitatively pointed out the limitation of the scalability in
TeraSort: excessive local disk I/O when the input gets enormous.
Though our analysis is based on the application of bioinformatics
which possess a characteristic of many small pieces of data,
we have deeply addressed how the scalability of TeraSort collapses
and found that keeping every suffix in place is the reason why 
TeraSort cannot scale well.

As such, keeping only the raw data in place has been proposed as a
conceptual guideline for our scheme to trade off memory I/O and network
against disk I/O.  Based on this guideline, we have proposed a scheme for 
scalable and efficient \ac{sa} construction built on \ac{mr} and distributed 
in-memory data store system.
Our scheme allows the suffixes to be generated from distributed in-memory data store system and
alleviates the self-expansion effect by taking the advantages of memory I/O and network.
Instead of passing the suffixes, \ac{mr} plays the role of communicating the 
indexes of suffixes to abate disk I/O during \ac{sa} construction.
In addition, we have developed our scheme with respect to load scalability,
structure scalability, and space scalability whereby our scheme can exhibit
better $scalability_1$.  Through the efficiency, we have shown that the $scalability_2$
of our scheme is better than TeraSort too.  All the experiments in the paper are
conducted using the authentic grouper genome and demonstrated to convince us that
our scheme can perform scalable and efficient \ac{sa} construction.

% conference papers do not normally have an appendix

% use section* for acknowledgement
%\section*{Acknowledgment}
%
%
%The authors would like to thank...
%more thanks here

% trigger a \newpage just before the given reference
% number - used to balance the columns on the last page
% adjust value as needed - may need to be readjusted if
% the document is modified later
%\IEEEtriggeratref{8}
% The "triggered" command can be changed if desired:
%\IEEEtriggercmd{\enlargethispage{-5in}}

% references section

% can use a bibliography generated by BibTeX as a .bbl file
% BibTeX documentation can be easily obtained at:
% http://www.ctan.org/tex-archive/biblio/bibtex/contrib/doc/
% The IEEEtran BibTeX style support page is at:
% http://www.michaelshell.org/tex/ieeetran/bibtex/
\bibliographystyle{IEEEtran}
{
\bibliography{ref}
}

\begin{acronym}
\acro{mr}[MR]{MapReduce}
\acro{sa}[SA]{Suffix Array}
\acro{gpus}[GPUs]{Graphics Processing Units}
\acro{hdfs}[HDFS]{Hadoop Distributed File System}
\acro{bwt}[BWT]{Burrows-Wheeler Transform}
\acro{gmm}[GMM]{Gaussian Mixture Model}
\acro{lcp}[LCP]{Longest Common Prefix}
\acro{mean}[$\mu$]{mean}
\acro{std}[$\sigma$]{standard deviation}
\acro{gc}[GC]{garbage collection}
\acro{am}[AM]{ApplicationMaster}
\end{acronym}

% that's all folks
\end{document}